# RADIAL MODE LITHIUM NIOBATE ROSEN TRANSFORMER

*Ziqian Yao[1], Heather Chang[2], Eric Stolt[2], Clarissa Daniel[2], Tzu-Hsuan Hsu[1], Juan Rivas-Davila[2], and Ruochen Lu[1]*
[1]The University of Texas at Austin, US, and [2]Stanford University, US

## ABSTRACT

In this work, we demonstrate the first two-port radial-mode Rosen transformer based on 36°Y-cut lithium niobate (LN) for piezoelectric power conversion. The device achieves a high transformation ratio (TF) of 16, a high electromechanical coupling factor ($k^2$) of 16.8% and a quality factor ($Q$) of 2500, yielding an outstanding figure of merit (FoM = $Q \cdot k^2$) of 420. The fabricated transformer features a large effective turns ratio of 16 and delivers an open-circuit voltage gain of 45.56 (unloaded) and 40.57 with a load of 1 MΩ in parallel with 0.1 pF, validating its ability to provide efficient passive voltage amplification. An equivalent circuit model was developed to accurately fit both finite element-simulated and measured admittance spectra, enabling reliable parameter extraction. These results establish LN radial-mode resonators as a promising high-performance, magnetic-less transformer platform.

## INTRODUCTION

As power systems strive for higher power density, magnetic transformers face increasing core losses, parasitic effects, and material constraints that limit further miniaturization [1]-[2]. These limitations have motivated the pursuit of alternative power conversion approaches that can achieve efficient voltage transformation within smaller footprints [3]-[4]. Piezoelectric transformers have emerged as a compelling candidate, as their ability to maintain high efficiency and power density in compact geometries makes them an attractive platform for next-generation compact power conversion systems [5]-[7].

Among various piezoelectric transformer configurations, Rosen-type transformers have been extensively studied for their ability to achieve large voltage step-up ratios and high passive voltage gain in compact architectures [8]-[9]. Early implementations using lead zirconate titanate (PZT) were widely adopted in power supplies and LCD backlighting systems [10]-[12]. The performance of PZT-based Rosen transformers, though effective at moderate power levels, was ultimately constrained by intrinsic material drawbacks, including high dielectric loss and thermal instability under large drive conditions, which degraded efficiency and long-term reliability [13]-[15]. As display technology transitioned to low-voltage LED lighting, the demand for high-voltage AC conversion decreased, resulting in a decline in research activity. More recently, piezoelectric transformers have regained attention as a promising solution for compact, high-efficiency, magnetic-less power conversion, driven by the growing need for integrated high-voltage amplification in modern electronics[16]-[17].

With advancements in piezoelectric materials, transformers using aluminum nitride (AlN) [18]-[19] and quartz [20]-[21] have demonstrated the capability of efficient operation at elevated frequencies. However, each material system presents intrinsic trade-offs that limit voltage amplification and energy transfer efficiency. AlN exhibits relatively low electromechanical coupling ($k^2$), which restricts voltage gain and power throughput, while quartz achieves extremely high mechanical quality factors ($Q$) but possesses extremely low coupling.

Lithium niobate (LN) has recently emerged as a promising platform for high-performance transformers due to its combination of strong coupling, low acoustic loss, and excellent thermal stability. LN transformers demonstrated efficient power transfer and high voltage step-up in both longitudinal-extensional (LE) and thickness-extensional (TE) configurations, establishing LN's potential for compact, high-frequency power conversion [22]-[24]. More recently, radial-mode LN transformers have demonstrated that in-plane vibration with a central nodal region can minimize anchor loss, decouple impedance from resonance frequency, and support scalable device geometries [25].

Building on these insights, this work presents a two-port radial-mode Rosen transformer on 36° Y-cut LN, achieving $k^2$ = 16.8%, $Q$ = 2500, and a figure of merit (FoM = $Q \times k^2$) of 420. The transformer delivers an open-circuit voltage gain of 45.6 V/V, demonstrating the strong potential of radial-mode LN structure for compact, high-efficiency, magnetic-less piezoelectric power transformers.

## DESIGN AND SIMULATION

The design of the two-port LN radial-mode Rosen transformer is illustrated in Fig. 1. The device employs a 1 mm thick, 36° Y-cut LN wafer with top and bottom electrodes arranged in concentric geometries, as shown in Fig. 1(a). The top view in Fig. 1(b) highlights the inner electrode (VIN$_1$) and the outer ring electrode (VIN$_2$), while the bottom ground (GND) electrode is shown in Fig. 1(c). This electrode configuration creates a large capacitance contrast between port 1 (VIN$_1$ and GND, thickness electrical field excitation via $e_{33}$) and port 2 (VIN$_2$ and GND, lateral electrical field excitation via $e_{31}$ and $e_{32}$), similar to longitudinal Rosen transformers [23], which establishes a strong impedance ratio that enables substantial passive voltage gain, analogous to the turns-ratio effect in magnetic transformers. The LN thickness and electrode geometry are carefully optimized to balance large impedance contrast while maintaining strong electromechanical coupling across the radial mode for both ports.

To determine the optimal cut and coupling orientation, Fig. 2 plots the calculated piezoelectric coefficients and effective in-plane coupling factors as a function of in-plane rotation of 36-Y LN. Although certain LN wafer cuts exhibit slightly higher piezoelectric coefficients, the 36° Y-cut is selected as it can provide simultaneously high $e_{31}$ and $e_{32}$, as well as $e_{33}$, enabling the two-port excitation with balanced multi-directional coupling. The data in Figs. 2(a)

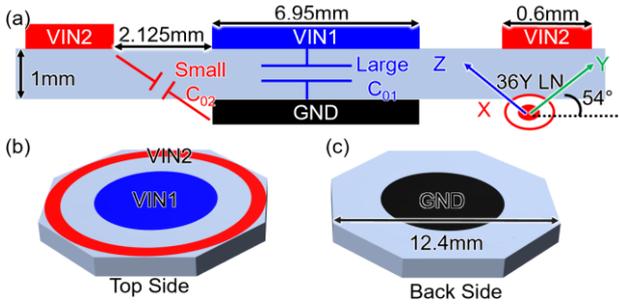

**Fig. 1** Schematics of (a) cross-sectional view with material axes. (b) Top and (c) back views.

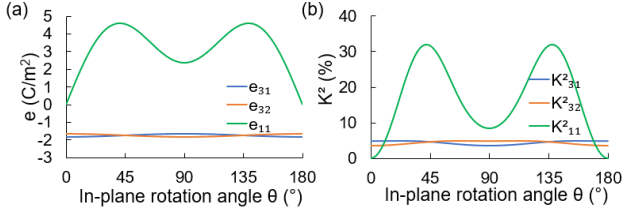

**Fig. 2** (a) Piezoelectric coefficients and (b) coupling coefficients versus in-plane rotation angle $\theta$ in 36°Y LN, highlighting high $e_{31}/e_{32}$ for radial mode excitation at Port 1, while high $e_{11}$ for harnessing charge at Port 2.

and 2(b) show the related piezoelectric coefficients, together with their associated coupling coefficients, enable efficient radial-mode operation for both ports.

To validate the proposed design, finite element analysis (FEA) was performed using COMSOL Multiphysics to simulate both the admittance spectra and the displacement field. Figure 3(a) shows the simulated and fitted admittance responses of the two-port transformer. Figs. 3(b) and 3(c) show two resonances observed near 313 kHz and 345 kHz, corresponding to the fundamental radial mode and a spurious higher-order wine-glass overtone, respectively, confirming effective mode excitation. As shown in Fig. 3(d), the radial mode exhibits concentric, axisymmetric expansion and contraction, with a zero-displacement node at the center of the disk, where the mounting is applied, and a maximum displacement near the outer edge. The corresponding lateral stress distribution in Fig. 3(e) further corroborates this behavior. The dominant stress component is confined to the active electrode region, with stress minimum located at the center of the disk. Still, the stress near the disk edge will be used for port 2 excitation.

To quantitatively model this behavior, an equivalent-circuit representation of the two-port radial-mode transformer is developed in Fig. 4, including both the main tone and the spurious mode. Each port is represented by a static capacitance $C_1$ and $C_2$. $C_3$ is the parasitic capacitance between ports. The acoustic resonance is captured by a motional branch comprising $R_m$, $L_m$, and $C_m$, representing mechanical loss, effective mass, and compliance, respectively. The motional branches are coupled through ideal transformers, which reflects the ratio of effective electrode areas and electromechanical transduction strength between the primary and secondary ports. The analytical expressions listed beneath Fig. 4 relate $R_m$, $L_m$, and $C_m$ to the

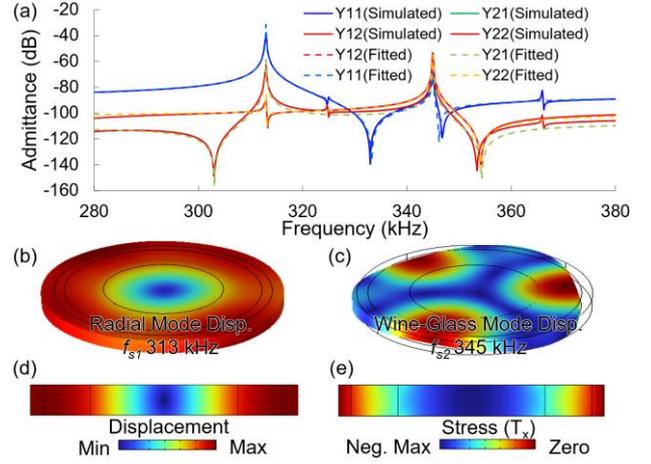

**Fig. 3** FEA simulation of Rosen transformer: (a) simulated and fitted admittance spectra, showing resonances at 313 kHz and 345 kHz, corresponding to (b) radial-mode and (c) wine-glass modes. (d) Cross-sectional displacement and (e) lateral stress $T_x$ for the radial mode.

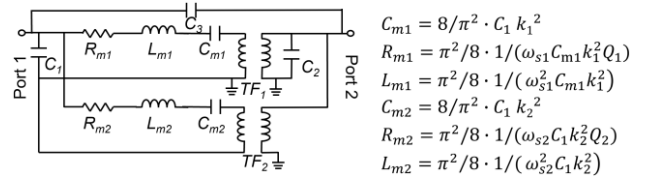

$$C_{m1} = 8/\pi^2 \cdot C_1 \, k_1^2$$
$$R_{m1} = \pi^2/8 \cdot 1/(\omega_{s1} C_{m1} k_1^2 Q_1)$$
$$L_{m1} = \pi^2/8 \cdot 1/(\omega_{s1}^2 C_{m1} k_1^2)$$
$$C_{m2} = 8/\pi^2 \cdot C_1 \, k_2^2$$
$$R_{m2} = \pi^2/8 \cdot 1/(\omega_{s2} C_1 k_2^2 Q_2)$$
$$L_{m2} = \pi^2/8 \cdot 1/(\omega_{s2}^2 C_1 k_2^2)$$

**Fig. 4** Equivalent circuit model for fitting of two-port Rosen transformer. Component values are defined by the listed equations.

measured resonant frequency $\omega_r$, coupling coefficient $k^2$, static capacitance, and mechanical quality factor $Q$. Fitting the simulated admittance spectra (plotted in Fig. 3) using this model allows for the accurate extraction of parameters. This model, therefore, provides a compact yet physically meaningful framework for correlating simulated and measured data, guiding the subsequent device fabrication and experimental validation.

## FABRICATION AND MEASUREMENT

The transformer was fabricated using standard microfabrication processes. The top electrodes (VIN$_1$ and VIN$_2$) were first defined through standard photolithography, followed by electron-beam metal deposition and lift-off to form the concentric input and output electrodes. A titanium (Ti) adhesion layer was deposited prior to the copper (Cu) layer to improve adhesion to the LN substrate. The total metal thickness was approximately 400 nm (Ti/Cu). The backside ground electrode (GND) was subsequently patterned using backside alignment to mirror the top VIN$_1$ region, ensuring proper field symmetry and uniform electric excitation across the LN thickness. Cu was chosen as the electrode material due to its high electrical conductivity, power-handling capability, and compatibility with subsequent wire-bonding and packaging processes. After metallization, the wafer was diced into individual dies as shown in Fig. 5(a).

For electrical characterization, the diced LN dies were mounted and wire-bonded onto a custom-designed PCB

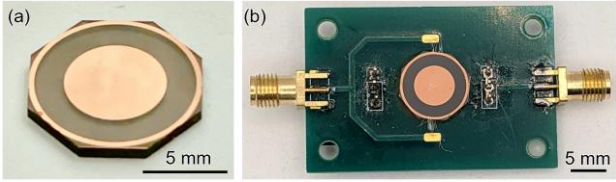

*Fig. 5* Fabricated device: (a) standalone element and (b) PCB-integrated two-port Rosen transformer.

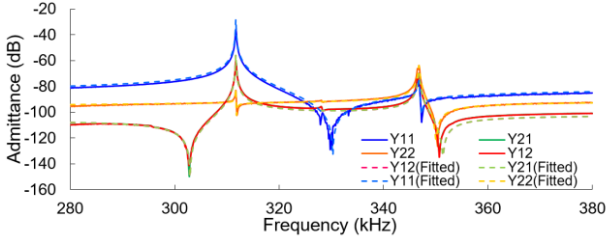

*Fig. 6* Measured and fitted admittance spectra of the two-port transformer. The extracted parameters, including resonance frequencies, coupling coefficients, quality factors, transformer ratios, and capacitances, are summarized in the table, which matches FEA results.

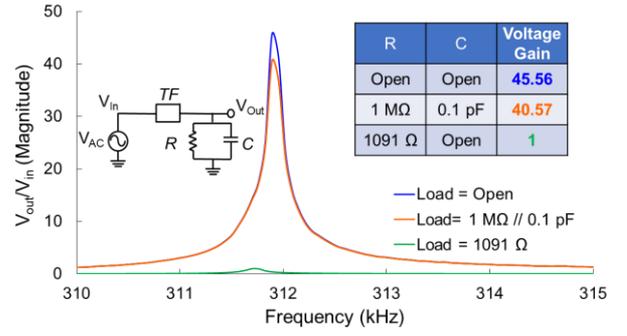

*Fig. 7* Circuit setup (left) and voltage gain in magnitude of the transformer under three load conditions: open circuit, 1 MΩ // 0.1 pF load modeling an active probe (GGB Model 12C), and 1091 Ω load giving unity gain. The table (right) lists the corresponding voltage gains.

| Reference | Platform | $f_s$ | $k^2$ | Q | FoM ($Q \cdot k^2$) | Load Condition | Voltage Gain |
|---|---|---|---|---|---|---|---|
| Lu, 2016 | AlN | 500 MHz | 1.24% | 1647 | 20.4 | Open | 7 |
| Manzaneque, 2017 | X-cut LN | 260 MHz | 28% | 480 | 134.4 | 50Ω | 14.4 |
| Galanko, 2019 | AT-cut Quartz | 49.8 MHz | 0.44% | 30200 | 132.8 | 1 MΩ // 0.1 pF | 28 |
| Phelps, 2024 | 128Y LN | 0.49 MHz | 2.8% | 4700 | 131.6 | 1 MΩ // 0.8 pF | 48.4 |
| **This work** | **36Y LN** | **0.31 MHz** | **16.8%** | **2500** | **420** | **1 MΩ // 0.1 pF** | **40.57** |

*Fig. 8* State-of-the-art comparison of two-port piezoelectric transformers.

using a suspended wire-bond integration scheme, as illustrated in Fig. 5(b). This approach mechanically decouples the resonator from the rigid PCB substrate, reducing contact stiffness and parasitic loading that would otherwise degrade the quality factor. The suspended wire bonds provide flexible electrical interconnects while maintaining low acoustic damping, ensuring that the measured response accurately reflects the intrinsic performance of the resonator.

The measured and fitted admittance spectra are presented in Fig. 6, exhibiting two distinct resonances corresponding to the radial and wine-glass modes. The extracted parameters yield a coupling coefficient of $k^2 = 16.8\%$, a mechanical quality factor of $Q = 2500$, and a FoM of 420, all in close agreement with simulation. The equivalent circuit fitting also provides an effective turns ratio of 16, along with the port capacitances and feedthrough capacitance, as summarized in Fig. 6.

The measurement setup and equivalent circuit were modeled and analyzed using Keysight Advanced Design System (ADS) to evaluate voltage amplification (Fig. 7). The two-port transformer was driven at Port 1 by an AC source, while the output voltage across Port 2 was monitored under different load conditions. The circuit model includes a parallel $R$–$C$ branch representing the external load, used to replicate various measurement environments. Three representative cases were examined: an open-circuit condition (no external load) to determine the intrinsic voltage gain; a 1 MΩ in parallel to 0.1 pF load, emulating the input impedance of an active RF probe (GGB Model 12C); and a 1091 Ω purely resistive load, corresponding to unity-gain power transfer. The voltage-gain spectra, shown in Fig. 7, demonstrate an open-circuit gain of 45.56 V/V, which decreases to 40.57 V/V under the probe-load condition and to unity under the 1091 Ω load. These results confirm the transformer's strong passive amplification capability and its predictable behavior under various loading conditions.

A state-of-the-art (SOA) comparison of two-port piezoelectric transformers is summarized in Fig. 8, benchmarking this work against previously reported devices. Earlier designs based on AlN and AT-cut quartz achieved either high $Q$ or moderate $k^2$, but their overall figures of merit (FoM = $Q \times k^2$) and voltage gains were limited by weak coupling or significant dielectric loss. X-cut LN devices demonstrated higher coupling (≈28%) but lower $Q$ values. More recent 128°Y-cut LN Rosen transformers have achieved voltage gains near 48 V/V with $Q \approx 4700$, representing the highest performance among prior LN-based structures. In comparison, the present 36°Y-cut LN radial-mode Rosen transformer achieves a balanced combination of high coupling ($k^2 = 16.8\%$), high quality factor ($Q = 2500$), and a record-high FoM of 420, all under a 1 MΩ ∥ 0.1 pF load. Despite operating at a lower frequency (≈0.31 MHz), the transformer delivers a substantial voltage gain of 40.57 V/V, validating the effectiveness of the radial-mode architecture. These results establish the proposed LN design as a high-performance platform that bridges the gap between high-$k^2$ AlN and high-Q quartz devices, offering superior passive voltage amplification for compact, magnet-less power conversion applications.

## CONCLUSION

In this work, a two-port radial-mode Rosen transformer based on 36°Y-cut LN was designed, fabricated, and experimentally demonstrated for efficient passive voltage amplification. The proposed device achieved a coupling coefficient of $k^2 = 16.8\%$, a mechanical quality factor of $Q = 2500$, and an outstanding FoM of 420, confirming excellent electromechanical performance. Finite element simulations and equivalent-circuit modeling showed strong agreement with measurements, validating the device design and analytical framework. Load-dependent measurements revealed an open-circuit voltage gain of 45.56 V/V, and 40.57 V/V under a 1 MΩ ∥ 0.1 pF load, demonstrating robust amplification capability. Compared with state-of-the-art piezoelectric transformers, this LN radial-mode architecture offers a balanced combination of high $k^2$, high $Q$, and low-frequency operation, enabling efficient, magnet-less power conversion for next-generation compact power electronics.

## ACKNOWLEDGEMENTS

This work was supported by the DARPA Nimble Ultrafast Microsystems (NIMBUS) program. The authors would like to thank Dr. Sunil Bhave for helpful discussions. Any opinions, findings, conclusions, or recommendations expressed in this material are those of the author(s) and do not necessarily reflect the views of the Defense Advanced Research Projects Agency (DARPA)

## CONTACT

*Z. Yao; hanson.yao@utexas.edu